\documentstyle[aps,preprint]{revtex}

\begin{document}

\preprint{May 23, 1997}

\title{
Heisenberg-picture approach to the evolution of the scalar fields
in an expanding  universe}
\author{K. H. Cho$^1$\footnote{E-mail: cho@thphy1.chonbuk.ac.kr},
J. Y. Ji$^2$\footnote{E-mail: jyji@phyb.snu.ac.kr},
S. P. Kim$^3$\footnote{E-mail: sangkim@knusun1.kunsan.ac.kr},
C. H. Lee$^4$\footnote{E-mail: chlee@hepth.hanyang.ac.kr},
J. Y. Ryu$^3$\footnote{E-mail: jyryu@knusun1.kunsan.ac.kr}}

\address{$^1$ Department of Physics, Chonbuk National University,
Chonju 561-756 ,   Korea\\
$^2$Department of Physics Education, Seoul National
University, Seoul 151-742, Korea\\
$^3$Department of Physics, Kunsan National University,
Kunsan 573-701, Korea\\
$^4$Department of Physics, Hanyang University,
Seoul 133-791, Korea}

\maketitle
\begin{abstract}
We present the Heisenberg-picture approach to
the quantum evolution of the scalar fields
in an expanding FRW universe which incorporates relatively
simply the initial quantum conditions such as the vacuum state,
the thermal equilibrium state, and the coherent state. We calculate
the Wightman function, two-point function, and correlation
function of a massive scalar field.
We find the quantum evolution of fluctuations of
a self-interacting field perturbatively and discuss
the renormalization of field equations.
\end{abstract}
\vspace{1cm}

\centerline{\it Submitted to Physical Review D as a Regular Article}

\newpage

\section{Introduction}

Quantum scalar fields such as the inflaton or Higgs fields in an expanding
universe are characterized by particle production
and {\it nonequilibrium} state during a rapidly expanding period.
Path integral and canonical method are two typical
methods to describe the quantum evolution and oftentimes to incorporate
an initial condition.  Path-integral formalism requires a modification
for the {\it nonequilibrium} process. The real-time formalism,
the imaginary-time formalism, and the complex-time formalism
are the field theoretic methods to describe such a {\it nonequilibrium}
quantum evolution of the field from an initial thermal equilibrium.
The real-time formalism has been frequently used to evaluate
the finite temperature Green's function in the expanding
FRW universe~\cite{niemi,semenoff,ringwald}. The canonical method
based on the Hamiltonian density operator
also has been widely used in which the functional Schr\"{o}dinger equation
describes the quantum evolution of the Higgs or scalar field in
the expanding universe~\cite{freese,eboli,guven}.
Functional Schr\"{o}dinger-picture has been
developed to find the wave function, bearing a close parallel to
the Schr\"{o}dinger-picture of quantum mechanics.

The purpose of this paper is to present the Heisenberg-picture approach
to describe the quantum evolution of the scalar fields in the
expanding FRW universe. The Heisenberg-picture has the advantage in
incorporating an arbitrary initial quantum condition in a particularly
simple form. For this purpose we reinterpret the quantum operators,
the so-called generalized invariants~\cite{lewis}, for a time-dependent quantum
system as constant operators in the Heisenberg-picture,
and in terms of which we find the quantum field in the Heisenberg-picture.
Extensive studies have already been done to find
and to make use of these invariants
for time-dependent oscillators. A remarkable point is that the exact
quantum states are the eigenstates of these invariant operators up to
time-dependent phase factors in the Schr\"{o}dinger-picture.
We applied the generalized invariants to the
Schr\"{o}dinger-picture description of the massive
scalar field in the expanding FRW universe~\cite{kim1,kim2} and Gao {\it et al}
also applied independently the Schr\"{o}dinger-picture to
both the massive scalar field and the self-interacting scalar field
in Ref.~\cite{gao}. Recently, the Heisenberg-picture, which has been
seldomly used, however, was employed to find the exact
quantum evolution for the same quantum systems~\cite{ji}.
Therefore, it would be worthy to apply the Heisenberg-picture approach
to the scalar fields in
the expanding FRW universe, which can be regarded as an infinite system of
decoupled time-dependent harmonic oscillators and
coupled time-dependent anharmonic
oscillators for the massive scalar field and the interacting scalar field,
respectively.

\section{Massive Scalar Field}

As the simplest model, we consider a massive scalar field
in the expanding FRW universe with the metric
\begin{equation}
ds^2 = - dt^2 + a(t)^2 d {\bf x}^2.
\end{equation}
The action is given by
\begin{equation}
{\cal S} =  \int  dt d^3 {\bf x} \frac{a^3}{2}
\Bigl[ \dot{\phi}^2 - \frac{1}{a^2} \bigl(\nabla \phi \bigr)^2
- \bigl( m^2 + \xi {\cal R} \bigr) \phi^2 \Bigr]
\end{equation}
where $\xi = 0$ for the minimal coupling and $\xi = \frac{1}{6}$
for the conformal coupling.
We Fourier-transform the scalar field
\begin{equation}
\phi ({\bf x}, t) = \frac{1}{(2\pi)^{3/2}}
\int d^3 {\bf k} \phi_{\bf k} (t) e^{i {\bf k} \cdot {\bf x}}
\end{equation}
and take the linear combinations
\begin{eqnarray}
\phi^{(+)}_{\bf k} (t) &=& \frac{1}{2}
\bigl(\phi_{\bf k} (t) + \phi_{-{\bf k}} (t) \bigr),
\nonumber\\
\phi^{(-)}_{\bf k} (t) &=& \frac{i}{2}
\bigl(\phi_{\bf k}(t) - \phi_{-{\bf k}} (t) \bigr).
\end{eqnarray}
Then the ${\bf x}$-integration yields
\begin{eqnarray}
\int d^3{\bf x} \phi^2 ({\bf x}, t)
&:=& \sum_{\alpha} \phi^2_{\alpha} (t)
=\int d^3 {\bf k} \phi_{\bf k} (t) \phi_{- {\bf k}} (t)
\nonumber\\
\int d^3 {\bf x} \pi^2 ({\bf x}, t)
&:=& \sum_{\alpha} \pi^2_{\alpha} (t)
= \int d^3 {\bf k} \pi_{\bf k} (t) \pi_{- {\bf k}} (t)
\end{eqnarray}
where $\alpha$ denotes $(\pm, {\bf k})$.
Here $\phi^{(+)}$ and $\phi^{(-)}$  should be treated as independent
variables corresponding to the cosine and sine modes in the box normalization.
Thus one obtains the Hamiltonian of mode-decomposed
time-dependent harmonic oscillators:
\begin{equation}
H(t) = \sum_{\alpha} H_{\alpha}
:= \sum_{\alpha} \Bigl[\frac{\pi_{\alpha}^2}{2 a^3(t)}
+ \frac{a^3 (t) \omega^2_{\alpha} (t)}{2} \phi^2_{\alpha}
\Bigr]
\end{equation}
where
\begin{equation}
\omega_{\alpha}^2 (t) = m^2 + \frac{{\bf k}^2}{a^2} +
\xi {\cal R},
\end{equation}
is the time-dependent frequency squared.
Due to the complete mode-decomposition,
the quantum states is a product of quantum state for each mode:
\begin{equation}
\vert \Psi (t) \rangle = \prod_{\alpha}
\vert \psi_{\alpha}(t) \rangle.
\end{equation}
Each quantum state obeys separately the Schr\"{o}dinger equation
\begin{equation}
i \frac{\partial }{\partial t} \vert \psi_{\alpha} (t)
\rangle = \hat{H}_{\alpha} (t) \vert \psi_{\alpha} (t) \rangle.
\end{equation}

As pioneered by Lewis and Riesenfeld~\cite{lewis},
the exact quantum states of a time-dependent quantum system can
be found in the Schr\"{o}dinger-picture as the eigenstates of
the generalized invariant operators.  We make use of these operators
as a method to find the quantum evolution of the scalar fields.
The invariant operator for the $\alpha$-th mode harmonic oscillator
satisfies
\begin{equation}
\frac{\partial}{\partial t} \hat{I}_{\alpha}
- i [ \hat{I}_{\alpha}, \hat{H}_{\alpha}] = 0.
\label{inv}
\end{equation}
There are many operators that satisfy Eq.~(\ref{inv}).
An invariant operator quadratic in momentum and position
was found explicitly for a time-dependent harmonic oscillator~\cite{lewis}.
In the Schr\"{o}dinger-picture, a pair of the invariant operators
that are linear in momentum and position, and can be interpreted as
the time-dependent creation and annihilation operators were found~\cite{kim1}
and were applied to the massive scalar field in the FRW universe~\cite{kim2}.
However, it is our observation that these invariant operators have a simple
interpretation in the Heisenberg-picture than in the Schr\"{o}dinger-picture.
We note that the time-dependent invariant operators in the
Schr\"{o}dinger-picture
\begin{eqnarray}
\hat{B}^{\dagger}_{\alpha} (t) &=&
-i[ \phi_{\alpha} (t) \hat{\pi}_{\alpha}
 - a^3 \dot{\phi}_{\alpha} (t) \hat{\phi}_{\alpha}],
\nonumber\\
\hat{B}_{\alpha} (t) &=& i[\phi^*_{\alpha} (t) \hat{\pi}_{\alpha} -
a^3 \dot{\phi}^*_{\alpha} (t) \hat{\phi}_{\alpha}] ,
\end{eqnarray}
where $\phi_{\alpha} (t)$ and $\phi^*_{\alpha} (t)$
are the complex solutions of the classical field equation
\begin{equation}
\ddot{\phi}_{\alpha} (t) + 3 \frac{\dot{a} (t)}{a(t)}
\dot{\phi}_{\alpha} (t) + \omega^2_{\alpha} (t) \phi_{\alpha} (t) = 0,
\end{equation}
can be interpreted as the time-independent operators
in the Heisenberg-picture
\begin{equation}
\frac{d}{dt} \hat{B}_{\alpha, H} (t) =
\frac{\partial}{\partial t} \hat{B}_{\alpha, H}
- i [ \hat{B}_{\alpha, H}, \hat{H}_{\alpha, H}] = 0.
\end{equation}
In order to have the usual commutation relation 
$[\hat{B}_{\alpha, H}, \hat{B}_{\alpha, H}^{\dagger}] = 1$,
the classical solutions satisfy the boundary condition
\begin{equation}
i a^3 \bigl( 
\dot{\phi}_{\alpha} (t) \phi^*_{\alpha} (t) 
- 
\phi_{\alpha} (t) \dot{\phi}^*_{\alpha} (t) 
\bigr) = 1.
\end{equation}
We denote these by $\hat{A}^{\dagger}_{\alpha}$
and $\hat{A}_{\alpha}$, respectively.
Thus, we find the  position and momentum operators in the Heisenberg-picture
\begin{eqnarray}
\hat{\phi}_{\alpha, H} (t) =
\phi_{\alpha} (t) \hat{A}_{\alpha}
+ \phi^*_{\alpha} (t) \hat{A}^{\dagger}_{\alpha} ,
\nonumber\\
\hat{\pi}_{\alpha, H} (t) =
a^3 [ \dot{\phi}_{\alpha} (t) \hat{A}_{\alpha}
+ \dot{\phi}^*_{\alpha} (t) \hat{A}^{\dagger}_{\alpha} ].
\end{eqnarray}
The Fock space of each mode is constructed from the vacuum state
\begin{equation}
\hat{A}_{\alpha} \vert 0_{\alpha}, t_0 \rangle = 0.
\end{equation}
The vacuum state of the scalar field is the product of the vacuum
state of each mode
\begin{equation}
\vert 0, t_0 \rangle = \prod_{\alpha} \vert 0_{\alpha}, t_0 \rangle.
\end{equation}

To apply this Heisenberg-picture to the massive Higgs field
and to incorporate the initial quantum conditions, we choose
$\hat{A}^{\dagger}_{\alpha}$ and {$\hat{A}_{\alpha}$ that
diagonalize the Hamiltonian at an initial time $t_0$
\begin{equation}
\hat{H}_{\alpha} (t_0) =
\omega_{\alpha} (t_0) \Bigl( \hat{A}^{\dagger}_{\alpha} \hat{A}_{\alpha}
+ \frac{1}{2} \Bigr).
\end{equation}
The expectation value of any operator in the Heisenberg-picture
is evaluated as
\begin{equation}
\langle \hat{O} \rangle = \langle \hat{O}_{H} \rangle_H.
\end{equation}
Below we consider separately three different types of initial conditions.

\subsection{Vacuum State}

Assuming that the initial quantum state be the $n$th number state
of the Hamiltonian, we find the Wightman function
of each mode
\begin{equation}
\langle n_{\alpha}, t_0 \vert  \hat{\phi}_{\alpha} (t)
\hat{\phi}_{\alpha} (t')
\vert n_{\alpha}, t_0 \rangle =
n_{\alpha} \Bigl( \phi_{\alpha}^* (t) \phi_{\alpha} (t')+
 \phi_{\alpha} (t) \phi_{\alpha}^* (t') \Bigr)
+ \phi_{\alpha} (t) \phi_{\alpha}^* (t').
\label{mode wightman}
\end{equation}
From Eq.~(\ref{mode wightman}) we find the Wightman function
for the scalar field
\begin{eqnarray}
\langle \hat{\phi} ({\bf x},t) \hat{\phi}({\bf x}', t') \rangle^{\rm (vac)}
&=&
\langle 0, t_0 \vert \hat{\phi} ({\bf x},t)
\hat{\phi}({\bf x}, t') \vert 0, t_0 \rangle
\nonumber\\
&=&
\frac{1}{(2 \pi)^3} \sum_{\alpha_1, \alpha_2} e^{i ({\bf k}_1 \cdot {\bf x}
- {\bf k}_2 \cdot {\bf x}')}
\langle \hat{\phi}_{\alpha_1} (t)
\hat{\phi}_{\alpha_2} (t') \rangle
\nonumber\\
&=&
\frac{1}{(2 \pi)^3} \int d^3 {\bf k} e^{i {\bf k} \cdot ( {\bf x}
- {\bf x}')}
{\phi}_{\bf k} (t)
{\phi}^*_{\bf k} (t')
\end{eqnarray}
The two-point function at an equal time follows
\begin{equation}
\langle \hat{\phi} ({\bf x},t) \hat{\phi}({\bf x}', t) \rangle^{\rm (vac)}
=
\frac{1}{(2 \pi)^3} \int d^3 {\bf k} e^{i {\bf k} \cdot ( {\bf x}
- {\bf x}')}
\vert {\phi}_{\bf k} (t) \vert^2,
\end{equation}
and so does the correlation function
\begin{equation}
\langle \hat{\phi}^2 ({\bf x},t) \rangle^{\rm (vac)}
=
\frac{1}{(2 \pi)^3} \int d^3 {\bf k} \vert {\phi}_{\bf k} (t) \vert^2.
\end{equation}

\subsection{Thermal Equilibrium}

Next we consider a thermal equilibrium as an initial state.
The initial thermal equilibrium is described by the density matrix
\begin{equation}
\hat{\rho}_{\alpha, I} (t_0 ) =
\frac{1}{Z_{\alpha, I}}e^{ - \beta \hat{H}_{\alpha} (t_0)},
\label{den op}
\end{equation}
where
\begin{eqnarray}
Z_{\alpha, I} &=& {\rm Tr} \bigl( e^{- \beta \hat{H}_{\alpha} (t_0)} \bigr)
\nonumber\\
&=& \Bigl( 2 \sinh \frac{\beta \omega_\alpha (t_0)}{2} \Bigr)^{-1}.
\end{eqnarray}
The expectation value  of  the operator is given by
\begin{eqnarray}
\langle \hat{O}_H \rangle^{\rm (therm)} &=& \frac{1}{Z_{\alpha,I}}
{\rm Tr} \bigl( e^{- \beta \hat{H}_{\alpha}
(t_0) } \hat{O}_H \bigr)
\nonumber\\
&=& \frac{1}{Z_{\alpha, I}} \sum_{n_{\alpha}}
 \langle n_{\alpha} \vert e^{- \beta \hat{H}_{\alpha}
(t_0) } \hat{O}_H \vert n_{\alpha} \rangle
\end{eqnarray}
It is straightforward to evaluate the Wightman function of each mode
\begin{eqnarray}
\langle \hat{\phi}_{\alpha} (t) \hat{\phi}_{\alpha} (t')
\rangle^{\rm (therm)} &=& \phi_{\alpha}^* (t) \phi_{\alpha} (t')
\frac{1}{Z_{\alpha, I}} \sum_{n_{\alpha}}
\langle \hat{A}^{\dagger}_{\alpha} \hat{A}_{\alpha} \rangle
e^{- \beta \omega_{\alpha} (t_0 ) \bigl( n_{\alpha} +
\frac{1}{2} \bigr)}
\nonumber\\
&& + \phi_{\alpha} (t) \phi_{\alpha}^* (t')
\frac{1}{Z_{\alpha, I}} \sum_{n_{\alpha}}
\langle \hat{A}_{\alpha} \hat{A}^{\dagger}_{\alpha} \rangle
e^{- \beta \omega_{\alpha} (t_0 ) \bigl( n_{\alpha} +
\frac{1}{2} \bigr)}
\nonumber\\
&=&
\frac{1}{ e^{\beta \omega_\alpha (t_0)} - 1}
\phi_{\alpha}^* (t) \phi_{\alpha} (t')
+ \frac{1}{ 1- e^{-\beta \omega_\alpha (t_0)}}
\phi_{\alpha} (t) \phi_{\alpha}^* (t').
\end{eqnarray}
The Wightman function of the scalar field is obtained by summing over modes
\begin{eqnarray}
\langle\hat{\phi}({\bf x}, t) \hat{\phi} ({\bf x}', t') \rangle^{\rm (therm)}
&=&  \frac{1}{(2\pi)^3}
\int d^3 {\bf k} e^{i {\bf k} \cdot ( {\bf x} - {\bf x}')}
\Bigl[ \phi_{\bf k}^* (t) \phi_{\bf k} (t')
\nonumber\\
&& + \frac{1}{e^{\beta \omega_{\alpha} (t_0)} - 1}
\Bigl(\phi_{\bf k}^*(t) \phi_{\bf k} (t')+
\phi_{\bf k} (t) \phi_{\bf k}^* (t') \Bigr)\Bigr]
\end{eqnarray}
The first term is the vacuum Green's function
and the second term is the thermal correction~\cite{semenoff}.
The two-point function and correlation function can be found
by taking the coincident limits $t = t'$ and $ {\bf x} = {\bf x}',
t = t'$, respectively.

\subsection{Coherent State}

As the last initial condition, we consider the coherent state
described by the density operator
\begin{equation}
\hat{\rho}_{\alpha, II} = \frac{1}{Z_{\alpha, II}}
e^{- \beta [ \omega_{\alpha} (t_0) \hat{A}^{\dagger}_{\alpha}
\hat{A}_{\alpha} + \gamma_{\alpha} \hat{A}^{\dagger}_{\alpha}
+ \gamma^*_{\alpha} \hat{A}_{\alpha} + \epsilon_\alpha ]},
\end{equation}
where
$ \epsilon_\alpha = | \gamma_\alpha |^2 /\omega_\alpha + \omega_\alpha /2 $.
The density operator can be unitarily transformed into the canonical form
(\ref{den op}) by acting the displacement operator
\begin{equation}
\hat{D}_{\alpha} =
e^{- \frac{\gamma_{\alpha}}{\omega_{\alpha} (t_0)} \hat{A}^{\dagger}_{\alpha}
 + \frac{\gamma^*_{\alpha}}{\omega_{\alpha}(t_0)}\hat{A}_{\alpha}}
\end{equation}
and so the partition function becomes
\begin{equation}
Z_{\alpha, II} = \frac{e^{- \beta \epsilon_\alpha}}
{1 - e^{- \beta {\omega}_{\alpha} (t_0)}}.
\end{equation}
We find the Wightman function of the scalar field
\begin{eqnarray}
\langle\hat{\phi}({\bf x}, t) \hat{\phi} ({\bf x}',t') \rangle^{\rm (coh)}
= && \frac{1}{(2\pi)^3}
\int d^3 {\bf k} e^{i {\bf k} \cdot ( {\bf x} - {\bf x}')}
\Bigl[
\phi_{\bf k} (t) \phi_{\bf k}^* (t')
+ \frac{1}{e^{\beta \omega_{\alpha} (t_0)} - 1}
\Bigl(\phi_{\bf k}^*(t) \phi_{\bf k} (t')
\nonumber\\
&& + \phi_{\bf k} (t) \phi_{\bf k}^* (t') \Bigr)
\Bigr]
+ \frac{1}{(2 \pi)^3} \int d^3 {\bf k}
\phi_{{\bf k}, c} (t) \phi_{{\bf k}, c}^* (t')
\end{eqnarray}
where
\begin{equation}
\phi_{{\bf k}, c} = \frac{\phi^*_{\bf k} (t) \gamma_{\bf k}
+ \phi_{\bf k} (t) \gamma_{\bf k}^* }{\omega_{\bf k}}
\end{equation}
is the classical field corresponding to the coherent state.

\section{Self-Interacting Scalar Field}

We now turn to the case of a self-interacting scalar field.
The Hamiltonian takes the form
\begin{equation}
H (t) =  \int  d^3 {\bf x}
\Bigl[ \frac{\pi_{\phi}^2}{2 a^3} + a^3 \Bigl(
\frac{1}{2a^2} \bigl(\nabla \phi \bigr)^2
+ V(\phi) \Bigr) \Bigr]
\end{equation}
We consider the potential of a general power law
\begin{equation}
V_1 (\phi) = \sum_n \frac{\lambda_n}{n!} \phi^n.
\end{equation}
and the potential of a particular cosmological interest 
\begin{equation}
V_2(\phi) = \frac{m^2}{2} \phi^2 + \frac{\xi {\cal R}}{2}
\phi^2 + \frac{\lambda}{4} \phi^4.
\end{equation}
Unlike the free massive scalar field, the Hamiltonian
cannot be simply decomposed as a sum over modes, but it is a coupled
system of modes. Thus, the canonical approach to the full quantum evolution
of the fields is extremely difficult to carry out explicitly.
However, we may find the quantum evolution of fluctuations around
the classical background field and include parts of the quantum back-reaction
of these fluctuations. By dividing the scalar field into the classical
background field and the fluctuations
\begin{equation}
\phi = \phi_c + \varphi
\end{equation}
such that
\begin{equation}
\langle \hat{\phi} \rangle = \phi_c (t),~ \langle \hat{\varphi} \rangle = 0,
\end{equation}
and by denoting $\pi_{\phi_c} = a^3 \dot{\phi}_c,
\pi_{\varphi} = a^3 \dot{\varphi}$,
we rewrite the Hamiltonian as
\begin{eqnarray}
H (t) = && \int  d^3 {\bf x}
\Bigl[ \Bigl( \frac{\pi_{\phi_c}^2}{2 a^3} +
\frac{a}{2} \bigl(\nabla \phi_c \bigr)^2
+ a^3 V(\phi_c) \Bigr)
\nonumber\\
&& + \Bigl( \frac{\pi_{\varphi}^2}{2 a^3} +
\frac{a}{2} \bigl(\nabla \varphi \bigr)^2
+ \frac{a^3}{2!} \frac{\delta^2  V(\phi_c) }{\delta \phi_c^2}
 \varphi^2 \Bigr)
\nonumber\\
&& + \Bigl( \frac{\pi_{\phi_c} \pi_{\varphi}}{a^3}
+ a \nabla \phi_c \cdot \nabla \varphi
+ a^3 \frac{\delta  V(\phi_c) }{\delta \phi_c} \varphi
+ \frac{a^3}{3!}  \frac{\delta^3  V(\phi_c) }{\delta \phi_c^3} \varphi^3
+ \frac{a^3}{4!} \frac{\delta^4  V(\phi_c) }{\delta \phi_c^4}
 \varphi^4 + \cdots \Bigr) \Bigr].
\label{int}
\end{eqnarray}
The canonical formalism is difficult to apply to the Hamiltonian (\ref{int}),
as a whole, consisted of the classical background field and
the fluctuations coupled each other. In this paper we treat the background
field $\phi_c$ as classical but evolve the fluctuations $\varphi$
quantum mechanically. It is, however, still difficult to proceed further
because we do not yet know the exact quantum states for the sub-Hamiltonian
of fluctuations, which is beyond quadratic.

As a constructive way to find the quantum evolution explicitly,
we truncate the Hamiltonian for the fluctuations at the quadratic order:
\begin{equation}
H_{(0)} (\varphi, t) = \int d {\bf x} \Bigl[
\frac{\pi_{\varphi}^2}{2 a^3} + a^3 \Bigl(
\frac{1}{2a^2} \bigl(\nabla \varphi \bigr)^2
+ \frac{1}{2!} \frac{\delta^2  V(\phi_c) }{\delta \phi_c^2}
 \varphi^2 \Bigr) \Bigr].
\label{trun ham}
\end{equation}
Then we evolve the full fluctuations perturbatively.
Then, as in the massive case, we decompose
the field by modes to the get the Hamiltonian as a
sum of time-dependent harmonic oscillators now with the frequency squared
\begin{equation}
\omega_{\alpha}^2 (t) = \frac{\delta^2  V(\phi_c) }{\delta \phi_c^2}
+ \frac{{\bf k}^2}{a^2}.
\end{equation}
The quantum back-reaction of the fluctuations
to the classical background field can be evaluated by taking
the quantum expectation values of fluctuation operators in the Hamiltonian:
\begin{eqnarray}
H (\phi_c, t) =  && \int  d^3 {\bf x}
\Bigl[ \frac{\pi_{\phi_c}^2}{2 a^3} + a^3 \Bigl(
\frac{1}{2a^2} \bigl(\nabla \phi_c \bigr)^2
+ V(\phi_c)
+ \frac{1}{2!} \frac{\delta^2  V(\phi_c)}{\delta \phi_c^2}
 \langle \hat{\varphi}^2 \rangle_{(0)}
 \nonumber\\
&& + \frac{1}{4!} \frac{\delta^4  V(\phi_c)}{\delta \phi_c^4}
 \langle \hat{\varphi}^4 \rangle_{(0)} + \cdots
 \Bigr) \Bigr].
\end{eqnarray}
In the above equation the expectation value of any odd power of the momentum
and field of quantum fluctuations such as $\langle \hat{\pi}_{\phi_c}
\hat{\pi}_{\phi} \rangle, \langle \nabla \hat{\phi}_c \cdot
\nabla \hat{\varphi} \rangle, \langle \hat{\varphi} \rangle,
\langle \hat{\varphi}^3 \rangle, \cdots$, vanishes for the initial
vacuum state and thermal equilibrium state. In these cases we see that
\begin{equation}
V_{eff} (\phi_c) =  V(\phi_c)
+ \frac{1}{2!} \frac{\delta^2  V(\phi_c)}{\delta \phi_c^2}
 \langle \hat{\varphi}^2 \rangle_{(0)}
+ \frac{1}{4!} \frac{\delta^4  V(\phi_c)}{\delta \phi_c^4}
 \langle \hat{\varphi}^4 \rangle_{(0)} + \cdots
\end{equation}
is an effective potential.
The Hamilton equations for the classical field equal to
\begin{equation}
\ddot{\phi}_c + 3 \frac{\dot{a}}{a} \dot{\phi}_c
+ \frac{\partial}{\partial \phi_c}
V_{eff} (\phi_c) = 0.
\label{clas fiel}
\end{equation}

\section{Renormalization}

It is to be noted that $\langle \hat{\varphi}^2 \rangle_{(0)}$,
$\langle \hat{\varphi^4} \rangle_{(0)}$, etc.,
contain infinite contributions and require
the renormalization of the coupling constants  $\lambda_n$.
Following Ref.~\cite{ringwald}, we do this by introducing
the counter-terms $\delta \lambda_n$
in order to cancel the infinite contributions
$\langle \hat{\varphi}^2 \rangle_{(0)}$,
$\langle \hat{\varphi^4} \rangle_{(0)}$, and etc..
The renormalized field equation for the classical background field is
\begin{eqnarray}
\ddot{\phi}_c + 3 \frac{\dot{a}}{a} \dot{\phi}_c
+ \Bigl( \sum_{n = 1} \frac{\lambda_{n}^{\rm (ren)}
+ \delta \lambda_n}{(n-1)!}
\phi_c^{n-2}
+
\sum_{n = 2} \frac{\lambda_{n}^{\rm (ren)} + \delta \lambda_n}{(n-2)!}
\phi_c^{n-3}  \langle \hat{\varphi}^2
\rangle_{(0)}
\nonumber\\
+
\sum_{n = 4} \frac{\lambda_{n}^{\rm (ren)} + \delta \lambda_n}{(n-4)!}
\phi_c^{n-5}  \langle \hat{\varphi}^4
\rangle_{(0)} + \cdots \Bigr) \phi_c = 0,
\label{ren clas}
\end{eqnarray}
where $\lambda_n^{\rm (ren)}$ are the renormalized coupling constants.
From the Appendix we find the expectation value of quantum fluctuations
with the initial vacuum state
\begin{equation}
\langle \hat{\varphi}^{2n} \rangle_{(0)}^{\rm (vac)} = \frac{(2n)!}{2^n n!}
\frac{1}{(2 \pi)^3} \int d^3 {\bf k} (\varphi^*_{\bf k} \varphi_{\bf k} )^{n}
\end{equation}
and with the initial thermal equilibrium state
\begin{equation}
\langle \hat{\varphi}^{2n} \rangle_{(0)}^{\rm (therm)} = \frac{(2n)!}{2^n n!}
\frac{1}{(2 \pi)^3} \int d^3 {\bf k} (\varphi^*_{\bf k} \varphi_{\bf k} )^{n}
\Bigl(1 +   \frac{2}{e^{\beta \omega_{\alpha} (t_0)} - 1} \Bigr),
\end{equation}
where each mode of the fluctuations obeys the classical equation
\begin{equation}
\ddot{\varphi}_{\alpha} + 3 \frac{\dot{a}}{a} \dot{\varphi}_{\alpha}
+ \Bigl( \frac{{\bf k}^2}{ a^2} + \sum_{n = 2}
\frac{\lambda_{n}^{\rm (ren)}}{(n-2)!}
\phi_c^{n-2} \Bigr) \varphi_{\alpha} = 0.
\label{ren fluc}
\end{equation}

We do explicitly for the potential $V_2$ by introducing the counter-terms
$\delta m^2, \delta \xi$ and $\delta \lambda$.
The renormalized field equation for the classical background field reads
\begin{eqnarray}
\ddot{\phi}_c && +  3 \frac{\dot{a}}{a} \dot{\phi}_c
+ \Bigl(m^{2{\rm (ren)}} + \delta m^2 + (\xi^{\rm (ren)}
+ \delta \xi) {\cal R}
+ ( \lambda^{\rm (ren)} + \delta \lambda) \phi_c^2
\nonumber\\
&& + 3 ( \lambda^{\rm (ren)} + \delta \lambda ) \langle \hat{\varphi}^2
\rangle_{(0)} \Bigr) \phi_c = 0.
\end{eqnarray}
The classical field equations for the fluctuations involve
only the renormalized coupling constants
\begin{equation}
\ddot{\varphi}_{\alpha} + 3 \frac{\dot{a}}{a} \dot{\varphi}_{\alpha}
+ \Bigl( \frac{{\bf k}^2}{ a^2} + m^{2{\rm (ren)}} + \xi^{\rm (ren)}
{\cal R}
+ 3 \lambda^{\rm (ren)} \phi_c^2 \Bigr) \varphi_{\alpha} = 0.
\end{equation}

The Wightman function of quantum fluctuations at tree level
can be found from the quantum evolution
of $\varphi$ using the truncated Hamiltonian (\ref{trun ham})
as in Sec. II. Higher order contributions to the Wightman
function can be found perturbatively by solving the classical background
field equation (\ref{ren clas}) and by substituting the classical background
field into the Hamiltonian (\ref{ren fluc}), and by repeating the procedure.

\section{Conclusion}

In this paper we presented the Heisenberg-picture approach to
the quantum evolution of scalar fields in an expanding FRW universe.
The quantum evolution of scalar fields is described by
the time-dependent functional Schr\"{o}dinger equation.
As a methodology to find the wave functions (quantum state)
of the functional Schr\"{o}dinger equation we made use of
the quantum operators called generalized invariants, which have
been used frequently in time-dependent quantum systems.
These generalized invariants have been applied mostly
in the Schr\"{o}dinger-picture.
Recently the Heisenberg-picture has been used to find the
quantum evolution of a time-dependent quantum system~\cite{ji}.

The mode-decomposed Hamiltonian of a massive scalar field in
the expanding FRW universe is a collection of time-dependent
harmonic oscillators. The generalized invariants that
may be used as the creation and annihilation operators
have been introduced~\cite{kim1} and used to find the
quantum evolution of the scalar field in the 
Schr\"{o}dinger-picture~\cite{kim2}. 
In the Schr\"{o}dinger-picture, however, it is rather
difficult to incorporate the initial quantum conditions.
It is our observation that it is relatively easy to
incorporate an arbitrary initial quantum condition
and that the generalized invariants have a simple interpretation
as constant operators in the Heisenberg-picture~\cite{ji}.
We found exactly the quantum evolution of the massive scalar field
incorporating the initial quantum conditions such as the
vacuum state, the thermal equilibrium state and the coherent state
in terms of the classical solutions and calculated the Wightman
function, the two-point function, and the correlation function incorporating
the same initial conditions. We also
found perturbatively the quantum evolution
of a self-interacting scalar field by dividing the scalar field
into a classical background field and fluctuations
and by handling the fluctuations quantum mechanically.
We found the renormalized field equation for the classical background
field equation.

\acknowledgments

We would like to thank Daniel Boyanovsky, Sung Ku Kim, Kwang-Sup Soh, and
Jae Hyung Yee for useful discussions.
KHC, SPK and JYR were supported by KOSEF under Grant No. 951-027-056-2,
SPK by BSRI Program under Project No. BSRI-96-2427, and CHL was supported
by KOSEF under Grant No. 94-0702-04-01-3.

\vskip 0.5 cm

\appendix
\section{Expectation Values}

In this appendix we evaluate the expectation values of the quantum
fluctuations with respect to the initial vacuum state and the initial
thermal equilibrium state.
Recollect the position operator of each mode of the scalar field
\begin{equation}
\hat{\varphi}_{\alpha} (t) = 
\varphi_\alpha (t) \hat{A}_{\alpha}
+ \varphi_\alpha^* (t) \hat{A}_{\alpha}^{\dagger}.
\end{equation}
We put the operator in the normal-ordered form 
\begin{eqnarray}
\hat{\varphi}_{\alpha}^{2n} = && 
\sum_{k} { {2n} \choose {2} } {{2n -2} \choose {2}}
\cdots {{2n - 2(k-1)} \choose {2}} \frac{1}{k!}
\bigl(\varphi_\alpha^* (t)  \varphi_\alpha (t) \bigr)^k
\nonumber\\
&& \times : \bigl(\varphi_\alpha (t)
\hat{A}_{\alpha} + \varphi_\alpha^* (t)
\hat{A}_{\alpha}^{\dagger} \bigr)^{2n - 2k}:
\end{eqnarray}
to get the vacuum expectation value
\begin{equation}
\langle \hat{\varphi}_\alpha^{2n} \rangle^{\rm (ren)}_{(0)}
= \frac{(2n)!}{2^n n!}
\bigl(\varphi_\alpha^* (t)  \varphi_\alpha (t) \bigr)^n .
\end{equation}

In order to evaluate the expectation value with respect to
the initial thermal equilibrium state, we make use of the following
theorem~\cite{louisell}
\begin{eqnarray}
\langle f (\hat{a}, \hat{a}^{\dagger}) \rangle^{\rm (therm)}
&=& \big(1 - e^{- \omega}\bigr)
{\rm Tr} f(\hat{a}, \hat{a}^{\dagger}) e^{- \omega \hat{a}^{\dagger}
\hat{a}}
\nonumber\\
&=&
\langle 0,0 \vert f[
\sqrt{1 + \bar{n}} \hat{a}
+ \sqrt{\bar{n}} \hat{c}^{\dagger},
\sqrt{1 + \bar{n}} \hat{a}^{\dagger}
+ \sqrt{\bar{n}} \hat{c}] \vert 0,0 \rangle
\end{eqnarray}
where
\begin{equation}
\bar{n} = \frac{1}{e^{\omega} - 1}
\end{equation}
and $\hat{c}$ and $\hat{c}^{\dagger}$ are the bosonic
operators which commute with $\hat{a}$ and $\hat{a}^{\dagger}$,
and $\vert 0, 0 \rangle$ is the vacuum state for $\hat{a}$
and $\hat{c}$.
Use the theorem to derive the expectation value
\begin{eqnarray}
\langle \hat{\varphi}_\alpha^{2n} \rangle^{\rm (therm)}
&=& \langle 0, 0 \vert
\Bigl[ \sqrt{ 1 + \bar{n}}
\bigl(
 \varphi_\alpha (t) \hat{A}_\alpha
 + \varphi_\alpha^* (t) \hat{A}_{\alpha}^{\dagger}
\bigr)
+ \sqrt{\bar{n}}
\bigl(
 \varphi_\alpha (t) \hat{C}_\alpha^{\dagger}
 + \varphi_\alpha^* (t) \hat{C}_{\alpha}
\bigr) \Bigr]^{2n}
\vert 0, 0 \rangle
\nonumber\\
&=& \sum_{k = 0}^{n} \langle 0, 0 \vert {{2n} \choose {2k}}
\Bigl[\sqrt{ 1 + \bar{n}}
\bigl(
 \varphi_\alpha (t) \hat{A}_\alpha
 + \varphi_\alpha^* (t) \hat{A}_{\alpha}^{\dagger}
\bigr) \Bigr]^{2k}
\nonumber\\
&& ~\times \Bigl[ \sqrt{\bar{n}}
\bigl(
 \varphi_\alpha (t) \hat{C}_\alpha^{\dagger}
 + \varphi_\alpha^* (t) \hat{C}_{\alpha}
\bigr) \Bigr]^{2n - 2k}
\vert 0, 0 \rangle
\nonumber\\
&=& \sum_{k = 0}^{n} {{2n} \choose {2k}}
\frac{(2k)!}{2^k k!}
\bigl(\sqrt{ 1 + \bar{n}} \bigr)^{2k} \bigl(\varphi_\alpha^* (t)
\varphi_\alpha(t)  \bigr)^{k}
\frac{(2n-2k)!}{2^{n-k} (n-k)!}
\bigl(\sqrt{\bar{n}} \bigr)^{2n - 2k}
\bigl(\varphi_\alpha^* (t) \varphi_\alpha(t) \bigr)^{n - k}
\nonumber\\
&=& \frac{(2n)!}{2^n n!}
\bigl(\varphi_\alpha^* (t) \varphi_\alpha(t) \bigr)^{n}
\sum_{k = 0}^{n} \frac{n!}{k! (n - k)!}
( 1 + \bar{n})^k \bar{n}^{n - k}
\nonumber\\
&=& \frac{(2n)!}{2^n n!}
\bigl(\varphi_\alpha^* (t) \varphi_\alpha(t) \bigr)^{n}
\bigl( 1 + 2 \bar{n} \bigr)^n,
\end{eqnarray}
where $\hat{C}_\alpha$ and $\hat{C}_\alpha^{\dagger}$ are the bosonic
operators which commute with $\hat{A}_\alpha$
and $\hat{A}_\alpha^{\dagger}$.

\end{document}